\documentclass[aps,prl,twocolumn,floatfix,superscriptaddress,amsmath,amssymb]{revtex4-1}

\usepackage{graphicx}
\usepackage{dcolumn}%
\usepackage{bm}
\usepackage[pdftex]{color}
\usepackage[dvipsnames]{xcolor}
\usepackage[colorlinks,bookmarks=false,citecolor=RoyalBlue,linkcolor=BrickRed,urlcolor=OliveGreen]{hyperref}
\usepackage[mathlines]{lineno}
\usepackage{braket}
\usepackage{bbm}
\usepackage{verbatim}
\usepackage{ulem}
\usepackage{standalone}

\newcommand{\parR}[1]{\noindent\textbf{\textit{#1}}\hspace{0.5cm}}

\usepackage{comment}

\usepackage[height=8.85in,width=6.45in]{geometry}
\usepackage{slashed}
\usepackage{braket}
\usepackage{tikz}
\usepackage{tikz-cd}
\usepackage{times}
\usepackage{courier}
\usepackage{bm}
\usepackage{xcolor}
\usepackage{mdframed}
\usepackage{datetime}
\usepackage{dsfont}
\usepackage{multirow}
\usepackage{cancel}
\usepackage{caption}
\usepackage{subcaption}
\usepackage{graphicx}
\usepackage[compat=1.1.0]{tikz-feynman}
\usepackage{pifont}

\newcommand{\KT}{\textbf{KT}}

\newcommand{\Z}{\mathbb{Z}}

\renewcommand\[{\begin{equation}}
\renewcommand\]{\end{equation}}

\def\ie{\begin{equation}\begin{aligned}}
\def\fe{\end{aligned}\end{equation}}

\begin{document}

\title{Noninvertible Symmetry-Enriched Quantum Critical Point}	

\newcommand{\ugent}[0]{Department of Physics and Astronomy, University of Ghent, 9000 Ghent, Belgium}
\newcommand{\ipmu}[0]{Kavli Institute for the Physics and Mathematics of the Universe, the University of Tokyo, 5-1-5 Kashiwanoha, Kashiwa, Chiba, 277-8583, Japan}

\author{Linhao Li}
\affiliation{\ugent}

\author{Rui-Zhen Huang}
\email{huangrzh@icloud.com}
\affiliation{\ugent}

\author{Weiguang Cao}
\email{weiguang.cao@ipmu.jp}
\affiliation{\ipmu}

\date{\today}

\begin{abstract}
Noninvertible symmetry generalizes traditional group symmetries, advancing our understanding of quantum matter, especially one-dimensional gapped quantum systems. In critical lattice models, it is usually realized as emergent symmetries in the corresponding low-energy conformal field theories. In this work, we study critical lattice models with the noninvertible Rep($D_8$) symmetry in one dimension. This leads us to a new class of quantum critical points (QCP), noninvertible symmetry-enriched QCPs, as a generalization of known group symmetry-enriched QCPs. They are realized as phase transitions between one noninvertible symmetry-protected topological (SPT) phase and another different one or spontaneous symmetry breaking (SSB) phase. We identify their low-energy properties and topological features through the Kennedy-Tasaki (KT) duality transformation. We argue that distinct noninvertible symmetry-enriched QCPs can not be smoothly connected without a phase transition or a multi-critical point.
\end{abstract}
\maketitle

\parR{Introduction.}
Recent studies of quantum matter have been significantly advanced by the exploration of  categorical symmetries $\mathcal{C}$, a generalization of traditional group-based symmetries \cite{Frohlich:2004ef,Frohlich:2006ch,Aasen:2016dop,Aasen:2020jwb,Bhardwaj:2017xup,Tachikawa:2017gyf,Chang:2018iay,Ji:2019jhk,Thorngren:2019iar,Thorngren:2021yso,Koide:2021zxj,Kaidi:2021xfk,Choi:2021kmx,Roumpedakis:2022aik,Bhardwaj:2022yxj,Cordova:2022ieu,Choi:2022jqy,Choi:2022zal,Kaidi:2022uux,Huang:2021zvu,Inamura:2021szw,Lootens:2021tet,Lootens:2022avn,Cao:2023doz,Li:2023ani,Seiberg:2023cdc,Seiberg:2024gek,Yan:2024eqz,cao2024generating,Okada:2024qmk,Gorantla:2024ocs,Pace:2024tgk,Choi:2023vgk}. Unlike conventional symmetries governed by group $G$ structures, $\mathcal{C}$ involve noninvertible symmetries which do not always permit inverse operations. 
Importantly, symmetry-protected topological (SPT) phases \cite{PhysRevB.85.075125,PhysRevB.87.155114,PhysRevB.83.035107,PhysRevB.80.155131}, originally proposed as nontrivial gapped phases with unique ground state protected by group $G$ symmetries, have been generalized to include symmetries associated with fusion categories $\mathcal{C}$, allowing for richer classifications of topological phases \cite{Thorngren:2019iar,Inamura:2021wuo,Cordova:2023bja,Bhardwaj:2023idu,Bhardwaj:2024qrf,Fechisin:2023dkj,Seifnashri:2024dsd}. 

A notable example is the Cluster SPT model
\ie \label{eq_ham_cluster}
H_{\text{cluster}} = \sum_j Z_{j-1}X_j Z_{j+1},
\fe
protected by $\Z_2 \times \Z_2$ group symmetry generated by even- and odd-site spin flip: $\eta^e=\prod_j X_{2j}$ and $\eta^o=\prod_j X_{2j-1}$.  It was recently found to have a larger Rep($D_8$) category symmetry~\cite{Seifnashri:2024dsd}. The symmetry operators correspond to irreducible representations (irreps) of the $D_8$ group, the dihedral group of order 8, and their product satisfies the same algebra as irreps of $D_8$. Specifically, the Rep($D_8$) is composed of invertible $\Z_2 \times \Z_2$ symmetry operators and a noninvertible symmetry operator \textbf{D}. As a symmetry, \textbf{D} commutes with the Cluster Hamiltonian $H_\text{cluster}$ following from its action 
\ie
&\textbf{D} \, X_j=Z_{j-1}Z_{j+1}\textbf{D},\quad  \textbf{D} \, Z_{j-1}Z_{j+1}=X_j\textbf{D}.
\fe
Furthermore, \textbf{D} satisfies a noninvertible fusion rule
\ie
\textbf{D}^2 = 1 + \eta^e + \eta^o + \eta^e\,\eta^o. 
\fe
There are three distinct Rep($D_8$) SPT phases. Besides the Cluster SPT, the other two SPTs, named Odd and Even SPT phases, are also proposed for Rep($D_8$) symmetric spin chains in Ref.~\cite{Seifnashri:2024dsd}. 

In critical lattice models, group symmetry-enriched quantum critical points (QCPs)~\cite{scaffidi2017gapless,10.21468/SciPostPhys.10.6.148,thorngren2021intrinsically,PhysRevX.11.041059,PhysRevLett.129.210601,yang2023duality,Li:2022jbf,wen2023bulk,Wen:2023otf,wang2023stability,Wen:2024udn,Huang:2024ror,Yu:2024toh,PhysRevB.109.245108} were proposed and classified as an analogy of gapped SPT phases. Nevertheless, the noninvertible symmetries are primarily realized as emergent phenomena within infrared (IR) conformal field theories (CFTs) at the QCP~\cite{Aasen:2016dop,Aasen:2020jwb,Bhardwaj:2017xup,Tachikawa:2017gyf,Chang:2018iay,Ji:2019jhk,Thorngren:2019iar,Thorngren:2021yso,Bhardwaj:2023bbf,Bhardwaj:2024qrf}. Although the generalized noninvertible symmetries are well-characterized in gapped systems including SPT phases and spontaneous symmetry-breaking (SSB) phases~{\cite{Bhardwaj:2023idu,Bhardwaj:2023fca,Bhardwaj:2024wlr,Bhardwaj:2024kvy,Chatterjee:2024ych}}, their role in gapless systems remains largely unexplored, raising the question of whether they can help our understanding of QCPs and phase transitions in such contexts. 

In this work, we 
construct lattice models that host noninvertible symmetry Rep($D_8$) in one dimension. Our results reveal the existence of a new class of critical states, termed \textit{noninvertible symmetry-enriched QCPs}, which generalize previously known group symmetry cases. These noninvertible symmetry-enriched QCPs are realized as quantum phase transitions (QPT) between one noninvertible SPT phase and either a different SPT or an SSB phase,  consistent with the expectation from field theories \cite{Bhardwaj:2024qrf,Bhardwaj:2024kvy}.

\noindent \parR{Kennedy-Tasaki duality transformation.} 
A universal and general theoretical framework for systematically identifying and classifying symmetry-enriched QCPs protected by $\mathcal{C}$ remains elusive. Notably, they usually involve complicated non-local string order parameters and boundary conditions. To uncover the properties of these QCPs, we leverage the Kennedy-Tasaki (\textbf{KT}) duality transformation. Here, before discussing noninvertible symmetry-enriched QCPs, we first show \textbf{KT}-duality as a simple and useful framework to construct lattice models for gapped noninvertible SPT phases.

\begin{figure*}[tbp]
\centering
\includegraphics[width=0.9\textwidth]{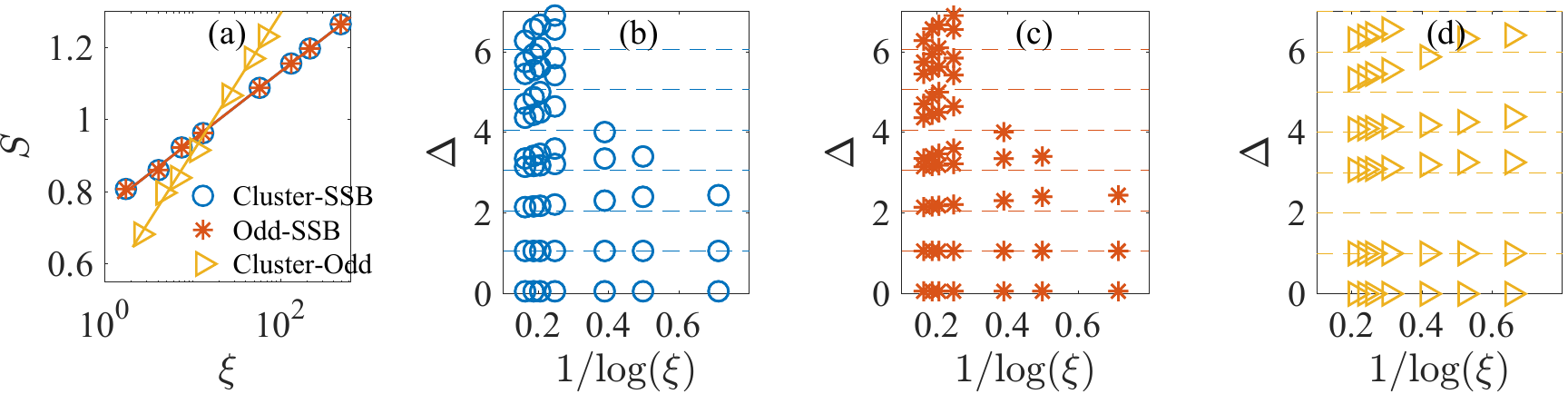}
  \caption{Universal entanglement properties in the ground states for the Rep($D_8$) QCPs obtained from variational iMPS. (a) Entanglement entropy vs finite correlation length $S \sim \frac{c}{6} \log (\xi)$. The central charge $c$ is fitted (line) to be $0.49$ for the Cluster-SSB and Odd-SSB QCPs, $1.00$ for the Cluster-Odd QCP. The $x$-axis is in logarithmic scale. (b-d) Rescaled and shifted low-level entanglement spectrum in the iMPS for the Cluster-SSB, Odd-SSB, and Cluster-Odd QCPs respectively. The first level is normalized to be $1/16$ in (b) and (c), or $0$ in (d).}
\label{fig_entanglement}
\end{figure*}

\textit{Noninvertible SPT phases constructed by} \textbf{KT}. The \textbf{KT} duality was introduced as a transformation that maps the SSB phase and  SPT phase of $\Z_2\times \Z_2$ spin chains and later generalized to general abelian groups \cite{1992CMaPh.147..431K, PhysRevB.45.304,oshikawa1992hidden,PhysRevB.78.094404,PhysRevB.54.4038,PhysRevB.88.125115,PhysRevB.88.085114,PhysRevB.83.104411,PhysRevB.107.125158,ParayilMana:2024txy}. It is also known as twisted gauging in the high energy literature \cite{Li:2023mmw}. Taking the Cluster SPT model $H_\text{cluster}$ as an example, \textbf{KT} maps it to a SSB phase 
\[
\begin{split}
\textbf{KT} \, Z_{j-1}X_jZ_{j+1} &= Z_{j-1}Z_{j+1}\,\textbf{KT},
\end{split}
\]
and vice versa. One can see it maps between symmetric operators representing different phases. 

More interestingly, considering a further $\Z_2$ symmetry $V := \prod_i \text{CZ}_{i,i+1}$, a product of controlled-Z gates on adjacent sites, \textbf{KT} can be used to construct $\mathcal{C}$-SPT models. In particular, it maps Rep($D_8$) symmetry quantum spin chain to a model with the anomalous $\Z^e_2\times \Z^o_2\times \Z^V_2$ symmetry~\cite{ref_SM}. All three Rep($D_8$)-SPT models can be constructed via the duality 
\ie\label{eq:KT-SPT model}
    G_0 \text{-SSB} + \Z^V_2\text{-SPT}
    &\stackrel{\textbf{KT}} \Longleftrightarrow   \text{Cluster-SPT} ,\\
    G_0\text{-SSB} + G_1\text{-SPT} &\stackrel{\textbf{KT}} \Longleftrightarrow   \text{Odd-SPT} ,\\
    G_0\text{-SSB} + G_2\text{-SPT} &\stackrel{\textbf{KT}} \Longleftrightarrow   \text{Even-SPT},
\fe
where $G_0 := \Z_2^{o} \times \Z_2^{e}$,  $G_1 := \text{diag}(\Z^V_2\times\Z^e_2)$ generated by $V\eta^e$, $G_2 := \text{diag}(\Z^V_2\times\Z^o_2)$ generated by $V\eta^o$, and the Odd SPT model takes the form \footnote{Here we add a factor 1/2 to the second and third term, compared to the Ref. \cite{Seifnashri:2024dsd}, which does not change the ground state.}
\ie \label{eq_ham_odd}
H_{\text{odd}}=\sum_{n} \, &Z_{2n-1}X_{2n}Z_{2n+1} - Y_{2n}X_{2n+1}Y_{2n+2}\,/\,2 + \\ 
&Z_{2n-1}Z_{2n}X_{2n+1}Z_{2n+2}Z_{2n+3}\,/\,2.
\fe
The Even SPT model $H_\text{even}$ is obtained by a one-site-shift of the $H_\text{odd}$. It is clear from the duality that distinct SPT phases can be interpreted as ``hidden symmetry breaking" with different unbroken symmetry patterns.

\textit{Group symmetry-enriched QCPs constructed by} \textbf{KT}.
The \textbf{KT} duality also provides a powerful tool for understanding and classifying group symmetry-enriched QCPs. As mentioned in the introduction, these critical states also enjoy topological features, since they involve further symmetries other than those shown in the IR CFT. Their construction and analysis are more complicated than gapped SPT phases due to the large symmetry in the corresponding IR CFT. Nevertheless, \textbf{KT} can construct symmetry-enriched QCPs from theories containing the decoupled low-energy CFT and gapped degrees of freedom (DOF). This approach allows for the convenient identification of topological features in QCPs enriched by different symmetry groups through a duality perspective, providing an interpretation through hidden symmetry breaking.  For example, from a $\Z^o_2$ SSB phase with a decoupled $\Z^e_2$ Ising CFT, the KT duality can yield a $\Z^o_2\times \Z^e_2$ enriched QCP as phase transitions between Cluster-SPT and $\Z^o_2$ SSB phase, which is known as the $\Z^o_2\times \Z^e_2$ gapless SPT (gSPT). A systematic analysis of group symmetry enriched QCP and gSPT was proposed~\cite{PhysRevX.11.041059,Li:2023knf}.

\parR{Rep($D_8$)-Enriched QCP.}
We are now prepared to introduce Rep($D_8$)-enriched QCPs. Two types of QCP can be realized in the Rep($D_8$) model, including QPTs between distinct SPT phases or between SPT and SSB phases. 

We first consider QCPs between different SPT phases, in which no symmetry breaking occurs. Here we focus on Cluster-Odd and Cluster-Even QCPs, as the Even-Odd QCP is a first-order discontinuous phase transition~\cite{ref_SM}. Such two QCPs are most conveniently understood from the dual viewpoint via the \textbf{KT}. The three Rep($D_8$)-SPT phases are dual to the different SSB phases, meanwhile the phase transition in the dual model is a ``deconfined quantum critical point" (DQCP) \cite{senthil2004deconfined,PhysRevB.70.144407}. Hence, the SPT-SPT QCP can be understood as a dual ``DQCP". 

\begin{table}[!tbp]
    \centering
    \begin{tabular}{|>{\centering\arraybackslash}p{1.5cm}|>{\centering\arraybackslash}p{1.2cm}|>{\centering\arraybackslash}p{1.0cm}|>{\centering\arraybackslash}p{1.0cm}|>{\centering\arraybackslash}p{1.0cm}|}
    \hline
       \multicolumn{2}{|c|}{} & Cluster & Odd  & Even \\
    \hline
       \multirow{2}{*}{PBC} & GSD    & 1 & 1 & 1 \\
                            & Charge & 2 & 2 & 2 \\
    \hline
       \multirow{2}{*}{$\Z_2^o$-TBC} & GSD    & 1 & 1 & 1 \\
                            & Charge & 2 & -2 & 2 \\
    \hline
       \multirow{2}{*}{$\Z_2^e$-TBC} & GSD    & 1 & 1 & 1 \\
                            & Charge & 2 & 2 & -2 \\
    \hline
        OBC & GSD & 4 & 4 & 4 \\
    \hline
    \end{tabular}
    \caption{Ground state degeneracy (GSD) and \textbf{D}-symmetry charges of Rep($D_8$)-SPT phases under PBC, TBC and OBC. The \textbf{D}-symmetry charge of Odd-SPT with $\mathbb Z_2^o$-TBC and that of Even-SPT with $\mathbb Z_2^e$-TBC distinguish them from the Cluster-SPT and each other. }
    \label{tab:SPTsignature}
\end{table}

To begin with, we study the QPT between the Cluster- and Odd-SPT phases. A similar analysis applies to the Cluster-Even QPT since the Even-SPT model is related to the Odd-SPT via a one-site shift. The QPT is realized by a single parameter interpolating between these phases
\ie
H = g \, H_{\text{cluster}} +  (1-g) \, H_{\text{odd}}
\fe
whose ground state is in the Cluster-SPT or the Odd-SPT phase for a small or large $g$, respectively. We are going to show a continuous  QCP exists at $g=1/2$ by using the \textbf{KT} duality.

By performing the duality transformation, the original Rep($D_8$)-model becomes
\[\label{eq:ham_cluster_odd}
\begin{split}
H_{\text{dual}}=& \,g \, H_\text{ssb}^1 + (1-g) \, H_{\text{ssb}}^2 ,
\end{split}
\]
in which the Cluster- and Odd-SPT are mapped to SSB models
\[
\begin{split}
H_{\text{ssb}}^1 = & \sum_{n} Z_{2n-1} Z_{2n+1} + Z_{2n} Z_{2n+2}, \\
H_{\text{ssb}}^2 = & \sum_{n} Z_{2n-1} Z_{2n+1} - Y_{2n}Y_{2n+2}(1+Z_{2n-1}Z_{2n+3})/2.
\end{split}
\]
One can find that the term $Z_{2n-1}Z_{2n+1}$ above commutes with the other terms in $H_{\text{dual}}$, hence at the low energy it is pinned down to be $Z_{2n-1}Z_{2n+1}=-1$, which means $\Z^o_2$ is always spontaneously broken. Hence the low-energy effective Hamiltonian is nothing but the XY-chain on the even sites
\[\label{eq:eff cluster-ood}
H_{\text{dual}}|_{\text{eff}} = g \, \sum_{n}Z_{2n-2}Z_{2n} - (1-g) \sum_n Y_{2n}Y_{2n+2}.
\] 
There is a direct and continuous critical point at $g=1/2$ from the antiferromagnetic $\mathbb Z_2^z$-SSB to ferromagnetic $\mathbb Z_2^y$-SSB. This effective model at the critical point is described by the compact boson CFT in the IR limit. It is easy to check that $\eta^e$ and $V$ acts as the $\pi$-rotation along $X$- and $Z$-direction in the effective model. As a result, there exists a QCP at $g=1/2$ between the Cluster-SPT and Odd-SPT phases whose bulk properties are described by a compact boson CFT.

Now we switch to studying the SPT-SSB QCPs. For lattice models protected by the categorical symmetry $\mathcal{C}$, the most interesting SSB phase is the one where ground states only break the noninvertible symmetry operator and preserve the invertible subcategory symmetry. In the Rep($D_8$) case, we denote such phase as the \textbf{D}-SSB phase, where ground states are invariant under invertible group symmetry $\Z_2\times \Z_2$ in Rep($D_8$) but break the noninvertible \textbf{D} symmetry \footnote{This SSB phase is also denoted as Rep($D_8$)/$\Z_2 \times \Z_2$ SSB in 
 Ref.~\cite{Bhardwaj:2024qrf}.}. 
A direct derivation of \textbf{D}-SSB is not easy. We again propose to use \textbf{KT}-duality to construct \textbf{D}-SSB spin chain. In particular, we use \textbf{KT} to map a $\Z^e_2\times \Z^o_2\times\Z^V_2$ symmetric chain in the $\Z^e_2\times \Z^o_2\times\Z^V_2$-SSB phase to a Rep($D_8$) model in the \textbf{D}-SSB phase
\ie
 \Z^e_2\times \Z^o_2\times\Z^V_2 \text{-SSB} &\stackrel{\textbf{KT}} \Longleftrightarrow   \textbf{D} \text{-SSB}.
\fe
A concrete lattice model $H_{\textbf{D}\text{-SSB}}$~\cite{ref_SM} can be readily obtained from \textbf{KT} and its ground states have two-fold degeneracy as a consequence of breaking \textbf{D} spontaneously. Using the ground states of the dual model and \textbf{KT}, the ground states of the \textbf{D}-SSB phase are spanned by
\ie\label{eq:wave D-SSB}
|\psi_1\rangle &= V[\otimes_{n}(|\rightarrow\rightarrow\rangle+i|\leftarrow\leftarrow\rangle)_{4n-3,4n-1}\otimes_n |\leftarrow\rangle_{2n}],\\
|\psi_2\rangle &= V[\otimes_{n}(|\rightarrow\rightarrow\rangle-i|\leftarrow\leftarrow\rangle)_{4n-3,4n-1}\otimes_n |\leftarrow\rangle_{2n}],
\fe
in which the symmetry properties can be examined explicitly.

Now let us discuss the QPT between the Rep($D_8$)-SPT and \textbf{D}-SSB phases. We start with the following Hamiltonian
\[ \label{eq:cluster-Dssb1}
\begin{split}
&H = g \, H_{\text{Rep($D_8$)-SPT}} + (1-g) \, H_{\textbf{D}\text{-SSB}},
\end{split}
\]
where the SPT model can be the Cluster- or the Odd-SPT model defined previously. To understand its low-energy properties and possible QPT, we perform an undoing-KT transformation to obtain its dual model. Despite the complicated interactions in Eq.~\eqref{eq:cluster-Dssb1}, its low-energy properties are surprisingly simple as revealed by the dual theory. The dual theory becomes effectively the conventional quantum Ising chain with $g$ being the transverse field~\cite{ref_SM}. We conclude there is a QCP at $g=1/2$ in the Rep($D_8$) model whose low-energy bulk properties are described by the Ising CFT. Or put another way, the Rep($D_8$)-QCPs are the KT dual of the conventional quantum Ising chain and additional high energy DOFs. Note that the QCPs between \textbf{D}-SSB and Cluster- or Odd-SPT belong to two distinct classes with very different topological features, which becomes clear in the next section. 

Besides the low-energy analysis in the dual model, the nature of these Rep($D_8$)-QCPs constructed above can be conveniently identified from finite entanglement scaling analysis and the universal entanglement spectrum in the ground state represented by infinite MPS (iMPS)~\cite{White:1992aa,pollman,Lauchli:2013aa,hrz1,hrz2,tns_review}. As shown in Fig.~\ref{fig_entanglement}, the Ising and compact boson CFTs are confirmed by the accurate central charge and universal entanglement spectrum. 

\parR{Topological features.}
Distinct SPT phases or symmetry-enriched QCPs are not smoothly connected, meaning that we can define topological features to characterize them. Again, we are going to rely on \textbf{KT} duality to find unique topological features that distinguish different classes. 

Firstly, similar to group SPT phases, one can utilize zero energy edge modes for open spin chains. Since the \textbf{KT} is an invertible transformation for open spin chains, the degeneracy or zero modes of Rep($D_8$) SPT with open boundary condition (OBC) can be easily found from the SSB phases in the dual model. In addition, the local order parameter in the dual model helps us design non-local string order parameters~\cite{ref_SM}. The most powerful and interesting topological feature is the noninvertible symmetry charge of the ground state under twisted boundary conditions (TBC) of invertible symmetry \cite{Li:2022jbf}.

We consider the spin chain which can be twisted by the $\Z_2^o\times\Z_2^e$ symmetry and label the state with $[(u_o,t_o),(u_e,t_e)]$, where $u$ and $t$ take values in $\{0,1\}$ denoting the $\Z_2$ charge and symmetry-twist, respectively. The  symmetry-twist is related with TBC as $Z_{2n+L}=(-1)^{t_e}Z_{2n}$, $Z_{2n-1+L}=(-1)^{t_o}Z_{2n-1}$. Denote the ground-state-energy for the original and the dual model in the sector $[(u_o,t_o),(u_e,t_e)]$ as  $E_{[(u_o,t_o),(u_e,t_e)]}$ and $\tilde{E}_{[(u_o,t_o),(u_e,t_e)]}$, which satisfy 
\begin{eqnarray}\label{eq:sptenergy}
    E_{[(u_o,t_o),(u_e,t_e)]}= \tilde{E}_{[(u_o, t_o+u_e),(u_e,t_e+u_o)]}.
\end{eqnarray}
In particular, consider the $\Z^o_2$-TBC, namely $t_o=1, t_e=0$.  
We have 
\begin{eqnarray}\label{eq:sptenergy}
    E_{[(u_o,1),(u_e,0)]}= \tilde{E}_{[(u_o, 1+u_e),(u_e,u_o)]} 
\end{eqnarray}
As dual theories of Rep($D_8$)-SPTs spontaneously break the $\Z^o_2\times \Z^e_2$ symmetry, the corresponding domain-wall costs finite energy. Hence they should have a trivial symmetry-twist in the ground state space, i.e., $1+u_e=u_o=0$. Thus the ground state of the Rep($D_8$)-SPT phase under $\Z_2^o$-TBC has charge $u_e=1, u_o=0$. This implies that Rep($D_8$)-SPT ground states under $\Z^o_2$-TBC are mapped to the ground states in the odd $\Z^e_2$ symmetry sector of SSB phase by \textbf{KT}. For the dual model (SSB) of the Odd-SPT, as $\text{diag}(\Z^V_2\times \Z^e_2)$ is unbroken, we can find the ground state under $\Z^o_2$-TBC $|\text{Odd}\rangle_{\Z^o_2\text{-TBC}}$ has a nontrivial \textbf{D} charge:
\[\label{eq:D charge}
\begin{split}
\textbf{D}\,|\text{Odd}\rangle_{\Z^o_2\text{-TBC}} &= \textbf{D} \, \text{KT}| \text{SSB}\rangle_{\left\{u_e=1, u_o=0\right\}}^{\text{dual}} \\
&= -2\,|\text{Odd}\rangle_{\Z^o_2\text{-TBC}}.
\end{split}
\]

In addition, for the dual models of Cluster-SPT and Even-SPT phase, as $\Z^V_2$ and $\text{diag}(\Z^V_2\times \Z^o_2)$ is unbroken repsectively, the ground states under $\Z^o_2$-TBC have trivial \textbf{D} charges. The result for all cases is summarized in Table~\ref{tab:SPTsignature}. 

\begin{table}[!tbp]
\centering
\begin{tabular}{|>{\centering\arraybackslash}p{1.1cm}|>{\centering\arraybackslash}p{1.0cm}|>{\centering\arraybackslash}p{0.9cm}| >{\centering\arraybackslash}p{0.9cm} | >{\centering\arraybackslash}p{0.9cm}| >{\centering\arraybackslash}p{0.9cm} | >{\centering\arraybackslash}p{0.9cm} |}
    \hline
         \multicolumn{2}{|c|}{} & SSB- & SSB-  & SSB- & Odd- &Even-  \\
         \multicolumn{2}{|c|}{} & Cluster & Odd & Even & Cluster &Cluster\\
    \hline
        \multirow{2}{*}{PBC} & GSD & 1 & 1 & 1 & 1 & 1 \\
                             & Charge & 2 & 2 & 2 & 2 & 2 \\
    \hline
        \multirow{2}{*}{$\Z_2^o$-TBC} & GSD & 1 & 1 & 1 & 2 & 1 \\
                             & Charge & 2 & -2 & 2 & $\pm$2 & 2 \\
    \hline
        \multirow{2}{*}{$\Z_2^e$-TBC} & GSD & 1 & 1 & 1 & 1 & 2 \\
                             & Charge & 2 & 2 & -2 & 2 & $\pm$ 2 \\
    \hline
         OBC & GSD & 4 & 4 & 4 & 2 & 2 \\
    \hline
\end{tabular}
\caption{Topological feature of symmetry-enriched QCPs: Ground state degeneracy (GSD) and \textbf{D}-symmetry charges of the ground state under PBC, TBC and OBC.}
\label{tab:gSPTsignature}
\end{table}

Now let us study the topological feature of Rep($D_8$) QCPs using the mapping of symmetry-twist sectors of the \textbf{KT} transformation. The topological feature of QPTs between Rep($D_8$)-SPTs and \textbf{D}-SSB is the same as the corresponding gapped SPT. However, for the Cluster-Odd and Cluster-Even QPTs, their topological properties differ from those of the gapped SPT phases. For instance, we can show the Cluster-Odd QCP under $\Z^o_2$-TBC ($t_o=1, t_e=0$) has two-fold degenerate ground states with \textbf{D} charges $\pm 2$ utilizing the \textbf{KT} duality. 

In the \textbf{KT}-dual model, the $\Z^o_2$ is spontaneously broken, which forbids the $\Z^o_2$ domain wall at the low-energy. Hence the ground state of Cluster-Odd QCP has odd $\Z_2^e$ charge, $u_e=1$. Then we focus on the ground-state energy in sector $[(u_o,1),(1,0)]$: $E^{\text{cluster-odd}}_{[(u_o,1),(1,0)]}=\tilde{E}^{\text{cluster-odd}}_{[(u_o, 0),(1,u_o)]}$. The dual theory at the low-energy is composed by the critical XY-chain and a gapped sector as analyzed above. This allows us to write down its ground state energy from the Bosonization analysis
\[
\tilde{E}^{\text{cluster-odd}}_{[(u_o, 0),(1,u_o)]} \sim \frac{\pi}{2L}[1 + (u_o)^2]+E_0.
\]
where the first and second term is contributed by the effective critical XY-chain and the gapped DOFs, respectively. Hence the ground states of Cluster-Odd QCP lie in the symmetry-twist sector $[(0,1),(1,0)]$. Moreover, the critical effective XY-chain ($g=1/2$) has two-fold degenerate ground states with nonzero magnetization $\langle\sum_n X_{2n}\rangle$ for $u_e=1$, as $\eta^e$ corresponds to the $\pi$-rotation along the $X$-direction, i.e., $\exp(i\pi \sum_n X_{2n})$. Thus $V$, effectively the $\pi$-rotation along $Z$-direction, exchanges the two ground states and two combinations of states carry even and odd $V$ charge. Finally, we obtain that the Cluster-Odd QCP has two-fold ground state degeneracy under $\Z^o_2$-TBC with \textbf{D} charges $\pm 2$. The results of \textbf{D} charge for other cases are summarized in Table~\ref{tab:gSPTsignature}.

\parR{Discussion and Outlook}
We have generalized group symmetry-enriched QCPs to noninvertible symmetries, specifically through one-dimensional lattice models with noninvertible Rep($D_8$) fusion category symmetry. 
These noninvertible symmetry-enriched QCPs exist as critical points between distinct phases, including noninvertible SPT phases and SSB phases, revealing new structural properties in quantum criticality. Our analysis of these QCPs’ low-energy behavior and topological properties via the \textbf{KT}-duality transformation has highlighted their unique role within the broader landscape of quantum states. Importantly, different classes of Rep($D_8$)-QCPs enjoy very different topological features, suggesting that noninvertible symmetry-enriched QCPs form a new class of states that cannot be smoothly connected without encountering either a phase transition or a multi-critical point. This insight suggests that noninvertible symmetries play an essential role in defining and classifying phase boundaries in gapless systems.

Looking ahead, our work opens several promising directions. First, further exploration of symmetry-enriched QCPs in higher-dimensional systems could deepen our understanding of these critical points’ generality and robustness across different spatial dimensions. In higher dimensions, unlike one-dimensional systems, the symmetries of the corresponding CFT are finite. And also it was already suggested that there exists more general symmetries than category symmetry. These difficulties make the analysis and understanding of higher dimensional QCPs much more challenging. Finally, developing experimental platforms to realize noninvertible symmetry-enriched QCPs, perhaps through quantum computers or cold-atom setups, would provide an opportunity to test our theoretical predictions and potentially uncover new phases of matter.

\parR{Acknowledgements.}
We thank Yunqin Zheng, Yuan Miao, Nai-Chao Hu and Jacob C Bridgeman for helpful discussions.

\bibliography{ref}

\newpage
\clearpage
\appendix
\widetext
\appendix

\section*{\centering Supplemental Material}

In this appendix, we show more technical details of the construction of the lattice models, the derivation of the topological features, and further numerical results.






\section{Properties of \textbf{D} and \textbf{KT} transformation}
In this section, we provide a brief review of various properties of the KT transformation, including the mapping between symmetry and twist sectors and the definition on open boundary conditions. The detail of  derivation is in Ref. \cite{Li:2023mmw,Seifnashri:2024dsd}.  

Let us consider a spin chain with $L\in 2\Z$ sites where each site hosts a spin-$\frac{1}{2}$, giving rise to a two-dimensional local Hilbert space spanned by the states $\ket{s_i}$ with $s_i=0,1$. The state can be acted upon by spin measurement and spin flip Pauli operators in the standard way, 
\begin{eqnarray}\label{eq:Paulioper}
Z_{i}\ket{s_i}= (-1)^{s_i} \ket{s_i}, \hspace{1cm} X_{i} \ket{s_i} = \ket{1-s_i}.
\end{eqnarray}
We also assume that the spin system has an on-site $\Z^o_2\times\Z^e_2$ global symmetry. States in this system can then be labeled as eigenstates of the operators $\eta^o$ and $\eta^e$ with the eigenvalues $(-1)^{u_o}$ and $(-1)^{u_e}$ respectively, where $(u_o,u_e)\in \{0,1\}^{\otimes 2}$.  Moreover, the $\Z^o_2\times\Z^e_2$ symmetry allows for twisting the boundary conditions as
\begin{equation}
\ket{s_{2i-1+L}}= \ket{s_{2i-1}+t_o}, \quad \ket{s_{2i+L}}= \ket{s_{2i}+t_e},
\label{eq:Ising_bc}
\end{equation}
where $(t_o,t_e)\in \{0,1\}^{\otimes 2}$. In summary, one can organize the Hilbert space into four symmetry-twist sectors, labeled by $[(u_o,t_o),(u_e,t_e)]\in \{0,1\}^{\otimes 4}$. 
In the condensed matter language, the symmetry-twist correspond to the TBCs:
\[
 Z_{2i+L}=(-1)^{t_e}Z_{2i}, \quad  Z_{2i-1+L}=(-1)^{t_o}Z_{2i-1}, \quad X_{i+L}=X_i.
\]

The theory after applying \textbf{D} or \textbf{KT} also possesses a $\Z^o_2\times\Z^e_2$ symmetry,  allowing symmetry-twist sectors to be defined, labeled by $[(\tilde{u}_o,\tilde{t}_o),(\tilde{u}_e,\tilde{t}_e)]\in \{0,1\}^{\otimes 4}$. 

Then \textbf{D} acts on the states as
\begin{eqnarray}
\textbf{D}\ket{\{s_j \}} = \frac{1}{2^{L/2}} \sum_{\{\tilde{s}_{j} \}} (-1)^{\sum_{j=1}^L s_{2j}(\tilde{s}_{2j-1}+ \tilde{s}_{2j+1}) + t_e \tilde{s}_{1} + \tilde{s}_{2j}(s_{2j-1}+ s_{2j+1}) + \tilde{t}_e s_{1} } \ket{\{\tilde{s}_{j}\}}.
\end{eqnarray}

The KT transformation is then defined as $\text{KT}=V\text{D}V$, which acts as 
\begin{eqnarray}\label{eq:ktdef2}
\begin{split}
    \text{KT} \ket{\{s_j \}} = \frac{1}{2^{L/2}} \sum_{\{\tilde{s}_{j} \}} (-1)^{\sum_{j=1}^L (s_{2j}+\tilde{s}_{2j})(\tilde{s}_{2j-1}+ \tilde{s}_{2j+1}+s_{2j-1}+ s_{2j+1}) + (t_e+\tilde{t}_e) (\tilde{s}_{1} +s_1)} \ket{\{\tilde{s}_{j}\}}.
\end{split}
\end{eqnarray}
The KT operator satisfies 
\[\label{eq:D-V relation}
\text{KT}\, \textbf{D}=2V\, \text{KT},
\]
hence the Rep($D_8$) symmetry is mapped to the $\Z^e_2\times \Z^o_2\times \Z^V_2$ symmetry with type III anomaly \cite{wang2015bosonic,PhysRevB.106.224420}.

Applying the KT transformation to the $\Z^o_2\times\Z^e_2$ symmetric model $H$ yields a dual model $\tilde{H}$ via $\tilde{H} \, \KT = \KT \, H$. Interestingly the symmetry-twisted sectors in the dual theory is closely related to the original one 
\begin{eqnarray}\label{eq:KTsectormaps}
[(\tilde{u}_o,\tilde{t}_o),(\tilde{u}_e,\tilde{t}_e)]= [(u_o, u_e+ t_o),(u_e, u_o+t_e)].
\end{eqnarray}
Here, the symmetry charges remain unchanged, but the boundary conditions are changed.  Even when there is no symmetry twist in the original theory, the boundary conditions in the dual theory may become twisted due to the symmetry charges of the original model. This yields the relation between $E_{[(u_o,t_o),(u_e,t_e)]}$ and $\tilde{E}_{[(u_o,t_o),(u_e,t_e)]}$, which is ground state energy of the original and the dual model in the sector $[(u_o,t_o),(u_e,t_e)]$ respectively:
\begin{eqnarray}\label{eq:sptenergy}
    E_{[(u_o,t_o),(u_e,t_e)]}=\tilde{E}_{[(\tilde{u}_o,\tilde{t}_o),(\tilde{u}_e,\tilde{t}_e)]}= \tilde{E}_{[(u_o, t_o+u_e),(u_e,t_e+u_o)]}.
\end{eqnarray}

Moreover, the KT transformation acts as a unitary transformation on an open interval, thereby preserving the energy spectrum of the model. This is particularly useful in identifying edge modes of gapped SPTs and symmetry-enriched QCPs.

\section{Topological feature of Rep($D_8$)-SPT}
In this appendix, we provide the detail on the topological feature of Rep($D_8$)-SPTs.
\paragraph{Edge modes under OBC:}
We begin by considering the Rep($D_8$) model with open boundary conditions (OBC). As analyized above, the dual models obtained by the KT transformation of the SPT models are in the SSB phases with four fold degeneracies. As KT now is a unitary transformation, we identify all of the three Rep($D_8$)-SPT phases are of four-fold degeneracy under OBC.
\paragraph{Symmetry charge under TBC of invertible symmetry:} Next, we consider the \textbf{D}-symmetry eigenvalue of different SPT ground states under $\Z^o_2$-TBC, namely $t_o=1, t_e=0$.  
Then we have 
\begin{eqnarray}\label{eq:sptenergy}
    E^{\text{SPT}}_{[(u_o,1),(u_e,0)]}= E^{\text{SSB}}_{[(u_o, 1+u_e),(u_e,u_o)]} 
\end{eqnarray}
As mentioned in the main text,  the ground state of Rep($D_8$)-SPT phase under $\Z^o_2$-TBC has charge $u_e=1, u_o=0$. This implies that the Rep($D_8$)-SPT ground states under $\Z^o_2$-TBC are mapped to the ground states in the odd $\Z^e_2$ symmetry sector of SSB phase by KT transformation. As  $\text{diag}(\Z^V_2\times \Z^e_2)$ is unbroken in the dual model of Odd SPT phase, following Eq.~\ref{eq:D-V relation}, we have
\[
\begin{split}
\textbf{D} |\text{Odd}\rangle_{\Z^o_2\text{-TBC}}& = \textbf{D} \, \text{KT}|\text{SSB-Odd}\rangle_{u_e=1, u_o=0} = 2 \, \text{KT} \, V|\text{SSB-Odd}\rangle_{u_e=1, u_o=0}\\
&=-2\text{KT}|\text{SSB-Odd}\rangle_{u_e=1, u_o=0}=-2 |\text{Odd}\rangle_{\Z^o_2\text{-TBC}}.
\end{split}
\]
Thus the ground state of Odd-SPT under  $\Z^o_2$-TBC carries a nontrivial charge of the noninvertible symmetry \textbf{D}. On the other hand, $\Z^V_2$ or $\text{diag}(\Z^V_2\times \Z^o_2)$ is unbroken for the SSB dual model of Cluster- or Even-SPT phase respectively. We have 
\[\label{eq:D charge}
\begin{split}
\textbf{D} \,|\text{Cluster}\rangle_{\Z^o_2\text{-TBC}}  = 2|\text{Cluster}\rangle_{\Z^o_2\text{-TBC}},\quad
\textbf{D} \, |\text{Even}\rangle_{\Z^o_2\text{-TBC}}& = 2 |\text{Even}\rangle_{\Z^o_2\text{-TBC}}.
\end{split}
\]
One can play the same game for all the other cases and the result is summarized in Table \ref{tab:SPTsignature}.

\paragraph{String order parameter:} 
At last, we discuss different string order parameters for the SPT phases which follow from the local order parameters for the dual SSB model under the KT transformation. The string order parameter can also be used to distinguish these three noninvertible SPT phases. For the Cluster phase, the long range order of the SSB dual model is given by the conventional local order parameter: 
\begin{eqnarray}
    m_{\text{SSB-Cluster}} = \braket{Z_{2i-1}} \braket{Z_{2i}}
\end{eqnarray}
For a finite system, the SSB can be detected by the two-point correlation functions and at large distance the correlation function approaches to the square of the order parameter defined above. Now one can apply KT to the correlation function of local operators to define a string order parameter
\begin{eqnarray}
    m_\text{Cluster} = \braket{Z_{2i-1} (\prod_{k=i}^{j-1}X_{2k} )Z_{2j-1}} \braket{Z_{2i} 
 (\prod_{k=i}^{j-1} X_{2k+1})Z_{2j}}.
\end{eqnarray}
These string order parameters also converge to non-zero finite values at large distance limit ($\vert i-j \vert \rightarrow \infty$), which serve as an order parameter for the Cluster-SPT.

Similarly, we can write down the string order parameter for the Odd-SPT phase. The long-range order in the dual SSB model of the Odd-SPT are characterized by 
\begin{eqnarray}
    m_\text{SSB-Odd} = \braket{Z_{2i-1}} \braket{O_i}
\end{eqnarray}
where $O_i := Y_{2i}(1-Z_{2i-1}Z_{2i+1})$. For the Odd-SPT phase, one can use the KT transformation of two-point correlations of the above local operator define a string order parameter
\ie
m_\text{Odd} = \braket{Z_{2i-1}\,\left(\prod_{k=i}^{j-1} X_{2k} \right)\,Z_{2j-1}} \, \braket{\tilde{O}_i\, \left(\prod^{j-1}_{k=i} X_{2k+1}\right) \, \tilde{O}_j}.
\fe
where $\tilde{O}_i:=(1-Z_{2i-1}X_{2i}Z_{2i+1})Y_{2i}$. We can also obtain the string order parameter of Even SPT phase, which is nothing but a one-site-shift of that for the Odd-SPT model.

\section{\textbf{D}-SSB phase}
In this section, we will discuss \textbf{D}-SSB phase in the Rep($D_8$) model, which is invariant under invertible $\Z_2\times \Z_2$ symmetry but breaks the noninvertible \textbf{D} symmetry operator. This phase is obtained by a KT duality transformation of a $\Z^e_2\times \Z^o_2\times\Z^V_2$-SSB phase. 

We start with a  $\Z^e_2\times \Z^o_2\times\Z^V_2$ symmetric lattice model whose ground state is in a fully SSB phase:
\[
\begin{split}
H_{\Z_2^{o} \times \Z_2^{e}\times \Z^V_2 \text{-SSB}}&=\sum_{n} Z_{2n-1}Z_{2n+1}+\sum_{n} Y_{4n-2}Z_{4n-1}Z_{4n+1}Y_{4n+2}\\&+\sum_n Z_{4n-3}Y_{4n-2}Y_{4n+2}Z_{4n+3}+\sum_n 2Z_{4n-4}Z_{4n-3}Z_{4n-1}Z_{4n}
\end{split}
\]
This model has eight SSB ground states which satisfy $Z_{2n-1}Z_{2n+1}=-1$, $Y_{4n-2}Y_{4n+2}=1$ and $Z_{4n-4}Z_{4n}=1$. These phases are characterized by a local order parameter
\ie \label{eq: LRE D8}
    m_\text{SSB} = \braket{Z_{2i-1}} \, \braket{Z_{4i}} \, \braket{Y_{4i-2}(1-Z_{4i-3}Z_{4i-1})Z_{4i}}
\fe
which breaks all the three $\Z_2$ symmetries. As analyzed above, the KT duality maps a $\Z^e_2\times \Z^o_2\times\Z^V_2$ symmetry to Rep($D_8$). In particular it maps the $\Z^e_2\times \Z^o_2\times\Z^V_2$-SSB phase to the Rep($D_8$) symmetric model with a \textbf{D}-SSB ground state
\[
\begin{split}
H_{\textbf{D}\text{-SSB}}&=\sum_{n} Z_{2n-1}X_{2n}Z_{2n+1}-\sum_{n} Y_{4n-2}Y_{4n-1}X_{4n}Y_{4n+1}Y_{4n+2}\\&-\sum_n X_{4n-1}X_{4n}X_{4n+1}Z_{4n-3}Z_{4n-2}Z_{4n+2}Z_{4n+3}-\sum_n 2Z_{4n-4}Y_{4n-3}X_{4n-2}Y_{4n-1}Z_{4n}
\end{split}
\]

The ground state preserving the invertible symmetry is as follows. As the first term and last term both commute with all other terms, the ground state satisfies $Z_{2n-1}X_{2n}Z_{2n+1}=-1$ and $Z_{4n-4}Y_{4n-3}X_{4n-2}Y_{4n-1}Z_{4n}=1$ at the low-energy. These two constraints give that the ground state satisfies $\prod_{n}X_{2n}=1$ and $\prod_n X_{2n-1}=1$, i.e., the invertible part of Rep($D_8$) is preserved in the ground state space. 

The degeneracy and the ground state wave function of the \textbf{D}-SSB model is not directly transparent. By performing the CZ-gate canonical transformation ($V$)~\footnote{Symmetry is not changed by the CZ transformation.}, the model can be mapped to
\[ \label{Eq:V_D_SSB}
\begin{split}
V H_{D\text{-SSB}}V^{-1} = &\sum_{n} X_{2n}+\sum_{n} (Z_{4n-3}X_{4n-2}Y_{4n-1}X_{4n}Y_{4n+1}X_{4n+2}Z_{4n+3}+\\
& Y_{4n-1}X_{4n}Y_{4n+1}Z_{4n-3}Z_{4n+3} + 2X_{4n-3}X_{4n-2}X_{4n-1})
\end{split}
\]
One can find all $X_{2n}$ and $X_{4n-3}X_{4n-1}$ commute with the Hamiltonian. It is clear that there are two SSB ground states. By acting $V$ onto ground states of the model Eq.~\eqref{Eq:V_D_SSB}, we now can write down the ground state of the \textbf{D}-SSB model as Eq.~\eqref{eq:wave D-SSB}.

Furthermore, the SSB of the non-invertible symmetry \textbf{D} can be characterized by a local order parameter, which is nothing but the third part of order parameter in \eqref{eq: LRE D8} after the KT transformation
\begin{equation}
      m_\text{D-SSB} = \braket{(1-Z_{4i-3}X_{4i-2}Z_{4i-1})Y_{4i-2}X_{4i-1}Z_{4i}}.
\end{equation}
Note that it is invariant under invertible $\Z^e_2\times \Z^o_2$ symmetry but Odd under \textbf{D} symmetry operator.

\section{QPT between Rep($D_8$) SPTs and SSB phase.}
In this appendix, we provide the detail of QPTs between Rep($D_8$) SPTs and \textbf{D}-SSB phase.

Let us start with the QPT between the Cluster-SPT and \textbf{D}-SSB phases. We consider the following Hamiltonian
\[ \label{eq:cluster-Dssb}
\begin{split}
&H_{\text{Cluster-SSB}} = g \, H_{\text{Cluster}} + (1-g) \, H_{\textbf{D}\text{-SSB}}
\end{split}
\]
in which a single parameter interpolates between different phases. The ground state is expected to host a continuous QPT between the Rep($D_8$)-SPT and \textbf{D}-SSB gapped phases. 

To understand its low-energy properties and possible QPT, we perform a undoing-KT transformation to obtain its dual model
\[
\begin{split}
H^\text{Cluster-SSB}_{\text{dual}}=&\sum_{n} Z_{2n-1}Z_{2n+1}+g\sum_n Z_{2n-2}Z_{2n}\\&+(1-g)\sum_{n} (Y_{4n-2}Z_{4n-1}Z_{4n+1}Y_{4n+2}+Z_{4n-3}Y_{4n-2}Y_{4n+2}Z_{4n+3}+2Z_{4n-4}Z_{4n-3}Z_{4n-1}Z_{4n})
\end{split}
\] 
As the first term and last term commutes with the other terms, the low energy state satisfies $Z_{2n-1}Z_{2n+1}=-1$ and $Z_{4n-4}Z_{4n}=1$ when $0<g<1$ which can be used to reduce the dual model. The spins at sites $4n$ are pinned down to all up or down. As a result, the Cluster model is dual to a trivial $\Z^V_2$-SPT and the \textbf{D}-SSB becomes a $\Z^V_2$-SSB, with a decoupled $\Z^o_2\times \Z^e_2$-SSB phase. The effective Hamiltonian for the dual model is nothing but the quantum Ising chain
\[
H^{\text{Cluster-SSB}}_{\text{dual}}|_{\text{eff}}=\pm 2\, g \sum_{n} Z_{4n-2} - 2(1-g)\sum_n Y_{4n-2}Y_{4n+2}
\]
where $\pm$ depends on the sign of spin state at $4n$ site. It is clear now the low-energy physics in the dual model at $g = 1/2$ is described by the 2d Ising CFT. Since the duality transformation does not change local dynamics, we conclude the Cluster-SSB QPT is realized in the model Eq.~\eqref{eq:cluster-Dssb} with $g = 1/2$. 

We further discuss the possible QPT between the \textbf{D}-SSB phase and Odd-SPT phase by considering the following Hamiltonian
\[ \label{Eq:odd-Dssb}
\begin{split}
H_{\text{Odd-SSB}} = g\,H_{\text{Odd}} + (1-g)H_{\textbf{D}\text{-SSB}}
\end{split}
\]
By undoing the \textbf{KT} transformation, we obtain a dual model
\[
\begin{split}
H^{\text{Odd-SSB}}_{\text{dual}}=&\sum_{n} Z_{2n-1}Z_{2n+1}-g\sum_n Y_{2n}Y_{2n+2}(1+Z_{2n-1}Z_{2n+3})/2\\&+(1-g)\sum_{n} (Y_{4n-2}Z_{4n-1}Z_{4n+1}Y_{4n+2}+Z_{4n-3}Y_{4n-2}Y_{4n+2}Z_{4n+3}+2Z_{4n-4}Z_{4n-3}Z_{4n-1}Z_{4n})
\end{split}
\]
As the first and third term commute with all other terms, the low energy state satisfies $Z_{2n-1}Z_{2n+1}=-1$ and $Y_{4n-2}Y_{4n+2}=1$. At the low-energy, all $4n$ sites are all pinned down to be in the same $Y$ eigen-state. Then the effective Hamiltonian again becomes the quantum Ising chain
\[
H^{\text{Odd-SSB}}_{\text{dual}}|_{\text{eff}}=\pm 2 g \sum_n Y_{4n} - 2(1-g)\sum_n Z_{4n-4}Z_{4n}
\]
Hence the Odd-SSB QPT is realized in by the model Eq.~\eqref{Eq:odd-Dssb} at $g=1/2$. At last, we point out that the Even-SSB QPT can be obtained by doing a 1-site translation for the Odd-SSB QPT.

One can see that the QPT between Rep($D_8$)-SPT and Rep($D_8$)-SSB are the KT version of the conventional quantum Ising chain and additional gapped sectors, or equivalently the undoing-KT of Rep($D_8$)-enriched QCPs is the tensor product of low-energy the quantum Ising chain and higher-energy gapped sectors. In the Rep($D_8$) model, the gapped sector of the dual model is coupled with the low-energy CFT fields. The topological features, e.g., the edge modes, the \textbf{D} charge under TBCs and string order parameter can be obtain by the same calculation as that of gapped SPT phases. Different classes of QPTs can not be smoothly connected with each other under local deformation..


\section{QPT between Rep($D_8$)-SPT phases}
In this appendix, we consider the phase transition between different Rep($D_8$)-SPT phases. In the whole phase diagram the Rep($D_8$) symmetry is preserved, including the QCPs.

\parR{Odd-Cluster} We start with phase transitions between the Odd SPT and Cluster SPT, i.e., $H_{\text{Odd-Cluster}}=\frac{1}{2} (H_{\text{Cluster}}+H_{\text{Odd}} )$
As mentioned in the main text, the KT-dual model of this Hamiltonian is 
\[
\begin{split}
H_{\mathrm{dual}}  = \sum_{n}Z_{2n-1}Z_{2n+1} + \frac{1}{2}\sum_n\, (Z_{2n-2}Z_{2n}- Y_{2n}Y_{2n+2}(1+Z_{2n-1}Z_{2n+3})
\end{split}
\]
and the low energy effective Hamiltonian is 
\[\label{eq:cluster-odd-dual}
H_{\mathrm{dual}}|_{\text{eff}}=\frac{1}{2}\sum_n (Z_{2n-2}Z_{2n}-Y_{2n}Y_{2n+2})
\]
As this effective Hamiltonian is described by the free boson CFT in the IR limit, it  has the following power-law decaying correlation:
\[
\begin{split}
  \braket{Z_{2i-1} Z_{2k+1}}=1, \braket{Z_{2i} Z_{2k+2}}\sim \frac{1}{|2k+2-2j|^{1/2}},\\ \braket{Y_{2i}(1-Z_{2i-1}Z_{2i+1}) Y_{2k}(1-Z_{2k-1}Z_{2k+1})}\sim \frac{1}{|2k-2j|^{1/2}}, 
  \end{split}
\] 
Thus, before KT transformation, the Cluster-Odd phase transition has the following string order 
\[
\begin{split}
  \braket{Z_{2i-1}\prod^k_{j=i} X_{2j} Z_{2k+1}}=1, \braket{Z_{2i} \prod^k_{j=i} X_{2j+1} Z_{2k+2}}\sim \frac{1}{|2k+2-2j|^{1/2}},\\ \braket{Y_{2i}(1-Z_{2i-1}X_{2i}Z_{2i+1})(\prod^{k-1}_{j=i} X_{2j+1}) Y_{2k}(1-Z_{2k-1}X_{2k}Z_{2k+1}))}\sim \frac{1}{|2k-2j|^{1/2}}, 
  \end{split}
\] 
As KT is unitary transformation on the open chain, the phase transition between Odd and Cluster SPT phases has two ground state degeneracy under OBCs. 

Then let us determine the $\Z^o_2$ charge of ground state, i.e., $u_o$. As discussed in the main text, we focus on the sector $[(u_o, 0),(1,u_o)]$ of the dual Hamiltonian \eqref{eq:cluster-odd-dual} and study its energy spectrum. As $\Z^o_2\text{-SSB}$ has two ground state degeneracy with $u_o=\pm 1$ and $\Z^e_2$ symmetry acts as the $\pi$ rotation along $X$ direction in the effective XY chain, we have
\begin{eqnarray}
    \tilde{E}^{\text{Cluster-Odd}}_{[(u_o,1),(1,u_o)]}=E^{\text{XY}}_{[(u_x=1,t_x=u_o)]}+E^{\Z^o_2\text{-SSB}}_{\text{gs}} 
\end{eqnarray}
where $E^{\text{effective}}_{[(u_x,t_x)]}$ is the ground state energy of symmetry-twist sector  $[(u_x,t_x)] \in \{0,1\}^{\otimes 2}$ and $E^{\Z^o_2\text{-SSB}}_{\text{gs}}$ is the ground state energy of $\Z^o_2\text{-SSB}$ phase.

By the bosonization approach, we can connect the spin operators and compact boson field in the low energy:
\[
Z_{2n}+iY_{2n}\sim e^{i\theta}(b_0(-1)^n+b_1 \sin\varphi), X_{2n}\sim \frac{1}{2\pi}\partial_x \varphi+(-1)^n \sin\varphi.
\]
Since $\eta^e$ and $V$ acts as the $\pi$ rotation along $X$ and $Z$ direction in this effective model, we know the they act on the low energy degree of freedom as:
\[
\eta^e: \theta\to \theta+\pi, \varphi\to\varphi,\quad V: \theta\to -\theta, \varphi\to-\varphi.
\]
The fields $\varphi(x,t)$ and $\theta(x,t)$ are subjected to the twisted boundary condition, 
\begin{equation}
    \varphi(x+L,t) = \varphi(x,t) + 2\pi m, \hspace{1cm} \theta(x+L,t)= \theta(x,t) + 2\pi n.
\end{equation}
After mode expansion, the fields are decomposed into zero modes and oscillator modes. Hence we have 
\begin{equation}\label{eq:modeexpansion}
    \varphi(x,t) \simeq 2\pi m \frac{x}{L} + \cdots, \hspace{1cm} \theta(x,t) \simeq 2\pi n \frac{x}{L} + \cdots 
\end{equation}
where $m,n$ are constrained by the twisted boundary conditions for $\varphi$ and $\theta$ respectively, which are determined by the charge and symmetry twists on the lattice. 
The ground state energy only receives a contribution from the zero modes, which gives $\frac{2\pi}{L}[\frac{1}{4}m^2 + n^2]$.

\begin{figure}[tbp]
\centering
\includegraphics[width=0.65\columnwidth]{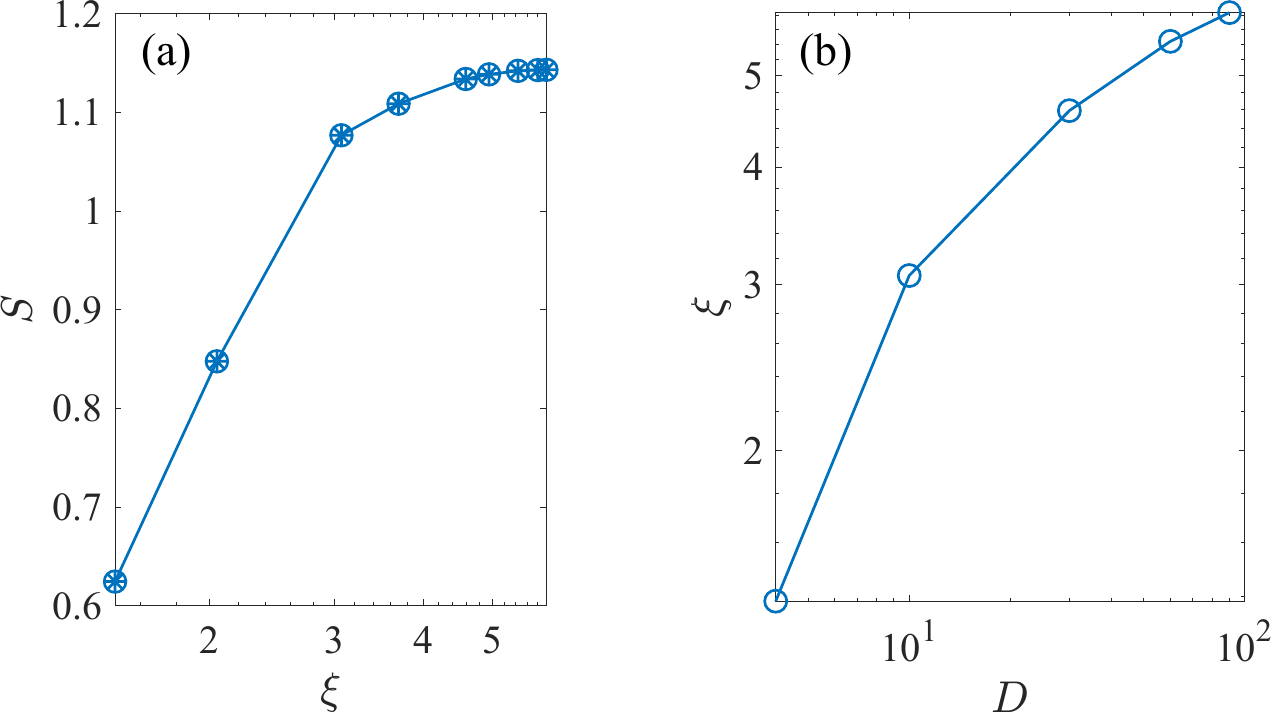}
  \caption{iMPS results for the Odd-Even QCP. (a) Entanglement entropy $S$ with respect to the correlation length $\xi$. The $x$-axis is in logarithmic scale. (b) The correlation length $\xi$ with respect to the bond dimension $D$ of the iMPS. The $x$- and $y$-axis are both in logarithmic scale.}
\label{fig_odd_even}
\end{figure}

As $\eta^e$, effectively the $\pi$ rotation along X direction, acts in the same way as $\exp(\int dx \frac{i}{2}\partial_x \varphi)=(-1)^m$, we have $u_x=m  ~ \text{(mod 2)}$. Moreover, there is also a relation between TBC  and boundary condition of $\theta$ as $t_x=2n ~\text{(mod 2)}$.  Thus ground state energy of effective dual model in $\Z^e_2$ symmetry-twist sector $(u_x=1,t_x=u_o)$   is 
\[
E^{\text{effective}}_{(1,u_o)}= \frac{\pi}{2L}[(1 + (u_o)^2].
\]
Moreover, the effective XY chain has two ground states in the sector $u_x=1, t_x=0$ corresponding to $m=\frac{1}{2\pi}\int\partial_x \varphi=\pm 1$. These two ground states indeed have the nonzero magnetization in the X-direction due to $m\sim \frac{1}{L}\sum_n X_{2n}$. As the $V$ maps $m$ to $-m$, $V$ exchanges these two ground states and we can have two combinations carrying even and odd $V$ charge. By performing the similar calculation in \eqref{eq:D charge}, we can obtain that the Cluster-Odd SPT transition has two ground state degeneracy under $\Z^o_2$ TBC, which lie in the symmetry-twist sector $[(0,1),(1,0)]$ and have $D$ charges $\pm 2$.

\parR{Even-Cluster} The Even-Cluster phase transition can be obtained by doing 1-site translation for the Even-Cluster phase transition. The above analysis applies directly.

\begin{figure}[tbp]
\centering
\includegraphics[width=0.42\columnwidth]{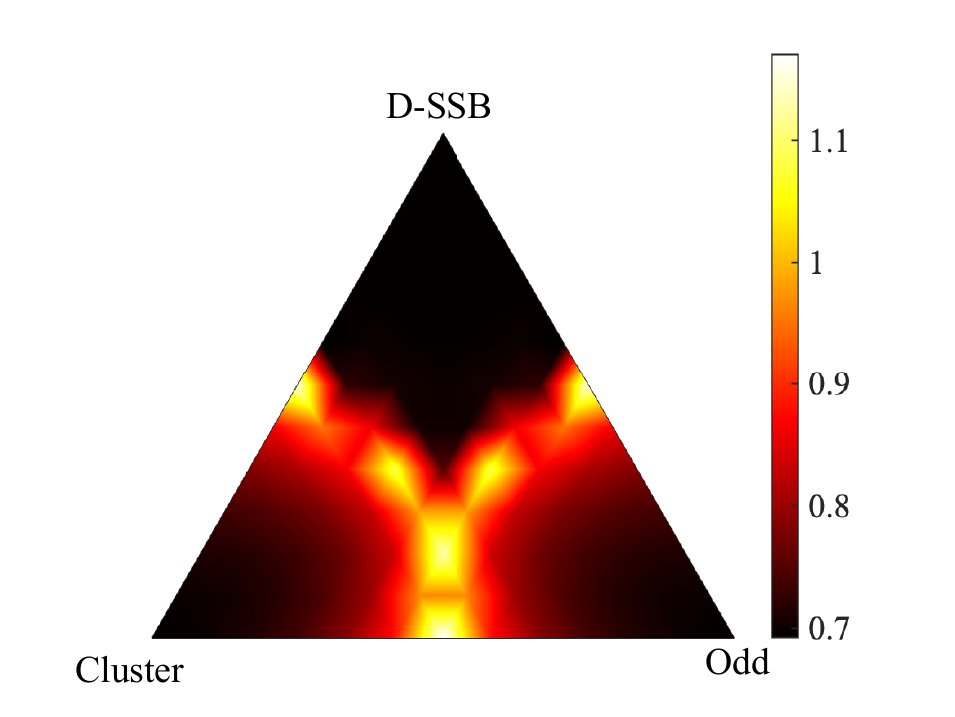}
  \caption{Ternary phase diagram of the Cluster-Odd-SSB model Eq.~\ref{H_abc}, where the color representing the entanglement entropy obtained from iMPS with a bond dimension $D=30$.}
\label{fig_abc}
\end{figure}

\parR{Cluster-Odd-SSB} Consider a model containing the Cluster, Odd and D-SSB phases
\ie \label{H_abc}
H = g_1\,H_\text{Cluster} + g_2\,H_\text{Odd} + g_3\,H_\text{D-SSB}
\fe
where $g_1 + g_2 + g_3 = 1$. The various QCPs belong to the middle point on the the axis, where one of the three parameter is zero. The nature of the QCP does not change when turning on the third parameter, untill the three critical lines merge to a multi-critical point. In Fig.~\ref{fig_abc}, we show the entanglement entropy and the numerical results support such a picture.

\parR{Odd-Even} We also studied the QPT between Odd- and Even-SPT phases
\ie
H_{\text{Odd-Even}} = g\,H_\text{Odd} + (1-g)\,H_\text{Even}.
\fe
As pointed out, the Odd and Even Hamiltonian are related by a 1-site translation, exchanging $g$ and $1-g$. A QPT is expected at $g = 1-g = 1/2$. The most likely picture, however, is a discontinuous transition resembling the QPT between different dimerized SSB phases.

The numerical iMPS results indeed support a first-order QPT, see Fig.~\ref{fig_odd_even}. The entanglement entropy $S$ becomes saturate to a constant value, ruling out the possibility of a finite central charge. Moreover, the length scale increase much slower than the power-law behavior and becomes convergent when increasing the bound dimension of iMPS. For a critical model described by a CFT, the correlation length becomes divergent and has a universal power-law form when increasing $D$~\cite{pollman}. 

\section{Fusion tensor of Rep($D_8$) symmetries in terms of MPO}
In this section, we study the fusion tensor which can be used to calculate the F-symbol of the Rep($D_8$) symmetry represented by matrix product operator (MPO). Note that the MPO symmetries studied in our work are realized in a physical lattice model in terms of local spins, different from those derived from anyon chain or strange-correlator in which the fusion tensor can be constructed directly. 

Assuming a fusion category $\mathcal{C}$ with a finite set of simple objects $\{O_i\}$, in which each of them is represented by a MPO with a finite $\chi$. Using the MPO framework, it is easy to calculate the deduction tensor and F-symbols of a fusion category thanks to the fundamental theorem of matrix product vectors. The product of two MPOs is also an MPO, $O_a \times O_b = O_{ab}$, which in general a sum of simple MPOs (simple object), $O_a\times O_b = N_{ab}^c O_c$. The virtual bond of MPO is nothing but the fusion space of the fusion category (possibly includes null vectors). There exists a deduction tensor, $X_{ab}^{c,\alpha}$ ($a,b,c$ labels of simple objects and $\alpha$ for the multiplicity), which can be used to extract these simple MPOs. In particular, the deduction tensor is nothing but canonical transformations on the virtual bond of the MPO. Once obtaining the fusion tensor $X^{ab}_{c,i}$, one can easily read the F-symbols
\ie
\includegraphics[scale=1.3]{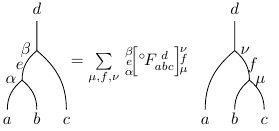}
\fe

All Rep($D_8$) symmetries can be efficiently represented by MP O with a finite virtual bond dimension $\chi$. The $\Z_2^e \times \Z_2^o$ invertible symmetries are MPOs with $\chi=1$, since they are tensor product of  on-site operators. The noninvertible symmetry \textbf{D} is the only nontrivial MPO with a bond dimension larger than one ($\chi = 4$). Note that the local tensor in an MPO can be conjugated by an arbitrary invertible transformation, though the MPO itself does not change. 

The fusion tensor involving trivial MPOs is simple. One only needs to find the fusion tensor for the $O_{\mathrm{\textbf{D},\textbf{D}}}^\textbf{D}$. The virtual space of the MPO $O_{\mathrm{\textbf{D,D}}}$ is isomorphic to the superposition that of all $\Z_2^e \times \Z_2^o$ MPOs which are all of one dimension. Note that there are $(\chi^2-4)$ null vectors in the whole $\chi^2$ space. In the canonical form, these null vectors can be identified by the zero entanglement spectrum. 

Let's first write down the local tensor for the \textbf{D}-MPO. As pointed out in Ref.~\cite{Seifnashri:2024dsd}, the noninvertible operator \textbf{D} is implemented as 
\begin{equation}
    \begin{aligned}
        \textbf{D} = T\,\textbf{D}^e\,\textbf{D}^o
    \end{aligned}
\end{equation}
where $T$ is the translation operator, $\textbf{D}^e$ and $\textbf{D}^o$ denote the Kramers-Wannier duality (\textbf{KW}) on the even and odd sites of the spin chain respectively. 
The local tensor for the MPO representation for the translation $T$ and \textbf{KW} takes the form of \textcolor{pink}{\cite{Seiberg:2024gek}}
\ie 
O_{T,j} = \left(
\begin{array}{cc}
    \frac{1+Z_j}{2}  &  \frac{X_j+iY_j}{2}\\
    \frac{X_j-iY_j}{2} & \frac{1-Z_j}{2} 
\end{array}
\right) \qquad \qquad  O_{\mathrm{\textbf{KW}},j} = \left(
\begin{array}{cc}
     H_j \frac{1+Z_j}{2} & H_j \frac{1+Z_j}{2} \\
     Z_j H_j \frac{1-Z_j}{2} & Z_j H_j \frac{1-Z_j}{2}
\end{array}\right)
\fe
where $H_j = \left(X_j + Z_j\right)/\sqrt{2}$ is the Hadamard quantum gate at site $j$.
From the expression of the local tensor in the $T$- and \textbf{KW}-MPO, one can easily construct the local tensor for the \textbf{D}-MPO with a $\chi=8$. After a gauge transformation on the virtual bond of MPO, one can find there are only $\chi=4$ non-zero space, and the local tensor can be written as
\ie 
O_{\mathrm{\textbf{D}},j} = \frac{1}{2\sqrt{2}}
\left(
\begin{array}{cccc}
 Z+1 & Z+1 & Z+1 & Z+1 \\
 X-i Y & X-i Y & -X+i Y & -X+i Y \\
 X+i Y & -X-i Y & X+i Y & -X-i Y \\
 1-Z & Z-1 & Z-1 & 1-Z \\
\end{array}
\right)_j
\fe
where the odd and even sites share the same form. One can easily verify the Rep($D_8$) fusion algebra using the explicit form of the MPOs.

The fusion tensor can also be written down as 
\ie
X_{\mathbf{\textbf{D}},\mathbf{\textbf{D}}}^a = \left(
\begin{array}{cccccccccccccccc}
 \frac{1}{2} & 0 & 0 & 0 & 0 & 0 & \frac{1}{2} & 0 & 0 & 0 & 0 & 0 & 0 & 0 & 0 & 0 \\
 -\frac{1}{2} & 0 & 0 & 0 & 0 & 0 & \frac{1}{2} & 0 & 0 & 0 & 0 & 0 & 0 & 0 & 0 & 0 \\
 0 & 0 & 0 & 0 & 0 & 0 & 0 & 0 & \frac{1}{2} & 0 & 0 & 0 & 0 & 0 & \frac{1}{2} & 0 \\
 0 & 0 & 0 & 0 & 0 & 0 & 0 & 0 & -\frac{1}{2} & 0 & 0 & 0 & 0 & 0 & \frac{1}{2} & 0 \\
\end{array}
\right)^t
\fe
where $a$ denotes the invertible symmetries and follow the order $\{I,\eta^e,\eta^o,\eta^e\eta^o\}$. The other fusion tensor for the fusion between \textbf{D} and $\Z_2\times \Z_2$ group elements are nothing but a gauging transformation for the \textbf{D}-MPO
\ie
X_{I,\mathbf{\textbf{D}}}^{\mathbf{\textbf{D}}} &= \left(
\begin{array}{cccc}
 1 & 0 & 0 & 0 \\
 0 & 1 & 0 & 0 \\
 0 & 0 & 1 & 0 \\
 0 & 0 & 0 & 1 \\
\end{array}
\right) \ \  
X_{\eta^e,\mathbf{\textbf{D}}}^{\mathbf{\textbf{D}}} = \left(
\begin{array}{cccc}
 1 & 0 & 0 & 0 \\
 0 & 1 & 0 & 0 \\
 0 & 0 & -1 & 0 \\
 0 & 0 & 0 & -1 \\
\end{array}
\right) \ \ 
X_{\eta^o,\mathbf{\textbf{D}}}^{\mathbf{\textbf{D}}} &= \left(
\begin{array}{cccc}
 0 & 1 & 0 & 0 \\
 1 & 0 & 0 & 0 \\
 0 & 0 & 0 & 1 \\
 0 & 0 & 1 & 0 \\
\end{array}
\right) \ \ \, \ \ 
X_{\eta^e\eta^o,\mathbf{\textbf{D}}}^{\mathbf{\textbf{D}}} = \left(
\begin{array}{cccc}
 0 & 1 & 0 & 0 \\
 1 & 0 & 0 & 0 \\
 0 & 0 & 0 & -1 \\
 0 & 0 & -1 & 0 \\
\end{array}
\right)
\fe
and
\ie
X_{\mathbf{\textbf{D}},I}^{\mathbf{\textbf{D}}} &= \left(
\begin{array}{cccc}
 1 & 0 & 0 & 0 \\
 0 & 1 & 0 & 0 \\
 0 & 0 & 1 & 0 \\
 0 & 0 & 0 & 1 \\
\end{array}
\right) \ \ \, \ \ 
X_{\mathbf{\textbf{D}},\eta^e}^{\mathbf{\textbf{D}}} = \left(
\begin{array}{cccc}
 1 & 0 & 0 & 0 \\
 0 & -1 & 0 & 0 \\
 0 & 0 & 1 & 0 \\
 0 & 0 & 0 & -1 \\
\end{array}
\right) \quad
X_{\mathbf{\textbf{D}},\eta^o}^{\mathbf{\textbf{D}}} &= \left(
\begin{array}{cccc}
 0 & 0 & 1 & 0 \\
 0 & 0 & 0 & 1 \\
 1 & 0 & 0 & 0 \\
 0 & 1 & 0 & 0 \\
\end{array}
\right) \ \ \, \ \ 
X_{\mathbf{\textbf{D}},\eta^e\eta^o}^{\mathbf{\textbf{D}}} = \left(
\begin{array}{cccc}
 0 & 0 & 1 & 0 \\
 0 & 0 & 0 & -1 \\
 1 & 0 & 0 & 0 \\
 0 & -1 & 0 & 0 \\
\end{array}
\right)
\fe
Now it is straightforward to use these deduction tensors above to show the F-symbol is indeed those for the Rep($D_8$) fusion category. 

\end{document}